\documentstyle[preprint,aps]{revtex}
\tighten
\begin{document}
\widetext
\preprint{NYU-TH/99/08/02}
\bigskip
\bigskip
\title{Constraints on Extra Time  Dimensions}
\medskip
\author{Gia Dvali$^{1}$, Gregory Gabadadze$^{1}$  and 
Goran Senjanovi\'c$^{2}$}

\bigskip
\address{$^1$Department of Physics, New York University, New York, NY 10003\\
$^2$International Center for Theoretical Physics, Trieste, 34014, Italy}
\bigskip
\medskip
\maketitle
\thispagestyle{empty}

\begin{abstract}
We discuss phenomenology of extra time
dimensions in a scenario where  the standard model particles are localized
in ``our'' time, whereas gravity can propagate in all time dimensions.
For an odd number of extra times, at
small distances, the (real part of the) Newtonian
potential is ``screened'' by  tachyonic Kaluza-Klein gravitons. In
general, the gravitational self-energy of objects acquires an imaginary
part. This complexity  may  either be interpreted as an amplitude for
the disappearance into ``nothing'', 
leading to the causality and probability
violation in low-energy processes, or as an artifact of the fictitious decay
into the unphysical negative energy tachyons.
The former case would put the most severe phenomenological restrictions on
such theories. In the latter case the size of extra times may be
within the reach of the proposed gravitational experiments.
In such a case these experiments should observe that the strength of the
Newtonian gravity diminishes  at short distances.

{}
      
\end{abstract}

\vspace{2in}
\begin{center}
{\it A contribution to the Yu. A. Golfand Memorial Volume}
\end{center}

 \narrowtext
\newpage

Extra space dimensions probed by
gravitational forces can be 
as large as a millimeter  without conflicting to   
any known laboratory or astrophysical bounds \cite{add}.
The extra dimensions which are usually considered 
are space-like. However, there is no  
{\it a priory} reason why extra times cannot exist. 
In this note
we will try to study some phenomenological constraints on such
hypothetical dimensions.
The main problem with time-like compactified dimensions from the
point of view of a low-energy field theory
is an existence of tachyonic modes.\footnote{The negative norm states
can also appear. These will not be discussed here. In our case
these modes have no direct couplings to matter. In some cases they
can be ``projected out'' if the translational invariance in the extra
time is broken. We thank Massimo Porrati for the discussions on related
issues.}
It is not clear whether 
an adequate quantum field-theoretical description of such states
can be formulated \cite {Nielsen}. 
This paper has no goal
whatsoever to deal with these problems. 
We will rather try to find obvious phenomenological 
constraints contradicting the existence of extra time-like
dimensions. We also suggest certain ideas which may weaken
these constraints.    

Consider for simplicity a five dimensional space-time
with a signature $(1,1,-1,-1,-1)$, with $\tau$ being an extra time
coordinate. 
In analogy with extra space coordinates one may
attempt to hide $\tau$  by compactifying it on a circle of
radius $L$. 
However, in contrast to the case of extra space-like dimensions
this does not do the job efficiently. 
The standard Kaluza-Klein excitations are now tachyonic states
with imaginary masses quantized in units of $i/L$.
In the space-like case, dimensions probed by standard model 
particles would be practically unobservable
for $L >$few TeV  or so \cite{alex}. However, the time-like case is very
different.
In fact, the 
exchange of KK states may lead to causality and probability
violations.
In \cite{ynd} it was argued that the probability violation essentially
rules out time-like dimension larger than the inverse Planck mass
$M_P^{-1}$. 
 The tree-level potential between two test charges mediated by the
exchange of a  photon propagating in one extra time has the form
\begin{equation}
V(r) \propto  { e^2 \over r}~\sum_{n}{\rm exp} 
\left(-in {r\over L}\right )~.  
\label{photon}
\end{equation}
This potential is complex.
The complexity of the amplitude may be interpreted in the way
that probability is not conserved
in the interactions  of two charged particles and they can escape into
``nothingness''.
The consideration of the baryon stability in the nuclei was argued to
set the bound $L \sim M_P^{-1}$ \cite{ynd}. 
Such a straightforward interpretation,
however, raises many questions. For instance, disappearance of the
particles would signal the charge and energy non-conservation,
although the starting theory is gauge and time-translationaly invariant.
Another important point is that the extra times may not be experienced
by the standard model interactions.

In the present paper we want to study constraints on
extra time dimensions,
in the light
of the possibility that the standard model particles can be
localized in the extra time  so that the corresponding KK modes 
are absent. We will also argue that complexity in the potential may have
a different interpretation, in which case the bounds would not be as severe.
We will be assuming that all the standard model particles
are localized at a particular time moment $\tau = 0$ and move freely only
in the remaining four coordinates $x_{\mu}, ~\mu=0,1,2,3$. 
They have localized
wave-functions
$\psi(x_{\mu})\delta(\tau)$ in  extra times. 
These particles can be
viewed as confined to 
a ``time-brane'', a 4-dimensional hyper-surface embedded in 
space-time with 
$q$ extra time dimensions of size L. 
The following toy example suggests a possible line of thought
how such a localization may come about. Consider a scalar field
$\phi$ in  space-time with one extra time-like dimension $\tau$.
Take the action
\begin{equation}
S_{2+3} = \int dx^{2+3} ~~~\left [(\partial_A\phi)^2 - V(\phi)\right ]~.
\end{equation}
Now choose the potential so that $\phi$ has a soliton-type solution
$\phi_{cl}(\tau)$ localized at say $\tau =0$ and independent of
$x_{\mu}$. This solution should satisfy the boundary 
conditions $\phi_{cl}(\pm \infty )={\rm const}$. 
If the approach to the asymptotic values is fast enough
there always exists  a normalizable zero mode
\begin{equation}
 \kappa = \partial_{\tau} \phi_{cl} ~g(x_{\mu})
\end{equation}
which is localized at $\tau = 0$.
From the four-dimensional 
perspective this may be viewed as a massless particle
satisfying the equation of motion
\begin{equation}
\partial^2_{3+1}~g(x_\mu)=0~.
\end{equation}
This state propagates  with the speed of light 
in our space-time, although
its wave-function in the extra time is 
peaked at $\tau=0$ moment and vanishes away from it.
This mode is a Goldstone boson
of spontaneously broken translation invariance 
in the $\tau$ direction and simply reflects
the fact that shifts of the soliton in $\tau$ does not cost any energy.
Similar considerations can be extended to states with different spins
using the line of arguments of Refs. \cite{ds}. 
This will not be attempted here.
Unfortunately, this method cannot completely get rid of tachyonic
states which can propagate in the extra time and also couple to the
ordinary matter. In particular such are KK gravitons (the ``time bulk
gravitons'')  which can
propagate in all space-time
dimensions and are viewed as tachyonic states by us.
On the other hand all the standard model particles can be localized
in the extra time.
Again, the above example should not be considered as a recipe 
for the realistic model building, instead it just  gives  
a crude idea of the localization in time. 
Below we will simply assume that the standard model  particles are
localized without further investigation of the precise mechanism
behind it. More details will be given elsewhere.
Let us compute the corrections to the gravitational energy due to
the existence of extra time dimensions. In the leading order these
arise due to the tree-level exchange of tachyonic KK states. 
Each
mode with mass $n/L$ generates a potential (for two test point masses $m$)
\begin{equation}
V_n(r) = G_N {m^2 \over r} {\rm exp}\left (-i n{r\over L} \right )~.
\end{equation}
The contribution to the self-energy  per unit mass of a gravitating body 
with an uniform density $\rho = M/R^3$ and size $R$ can be estimated as
\begin{equation}
{E \over M} =  4\pi G_N \int_0^R dr~r \rho \sum_{n=0}^{n_{max}}
{\rm exp} \left (-i{rn \over L} \right )~,
\end{equation}
where for $q$ extra times $n = \sqrt{n_1^2 +..+n_q^2}$ and summation
goes over all $n$'s up to a maximal value, which we will take as
$n_{max} \sim (M_{Pf}L)^N \sim (M_P/M_{Pf})^2$ (here 
$M_{Pf}$ is the fundamental Planck scale \cite {add}). 
For small $L$ one can
expand each integral in a series of $Ln/R$. In the leading order
this gives the following corrections to the real and 
imaginary parts of the energy
\begin{equation}
{ {\rm Re} E \over M} \sim 4\pi G_N {M L^2\over R^3}
\sum_{n=1}^{n_{max}}{1 \over n^2}~,
\label{self1}
\end{equation}
and 
\begin{equation}
{{\rm Im} E\over M} \sim 4\pi G_N{ M L^3\over R^4}\sum_{n=1}^
{n_{max}}{1 \over n^3}~.
\label{self2}
\end{equation}
The expression (\ref {self1}) is a correction to the 
gravitational self-energy
and is negligible for $L$  even as large as 1 millimeter 
(this is true even for a neutron star\cite{add}).

However, the physical meaning of expression (\ref {self2}) 
is somewhat less clear. Under the normal circumstances the
complexity of the self-energy would signal a non-zero decay
amplitude. In the present context, however, there is no obvious candidate 
to decay into (note that the tachyons with negative or imaginary energy
should be
regarded as unphysical, see below).
Thus, one option is that (\ref {self2}) signals that
probability is not conserved and
can be interpreted as a decay width into ``nothing''.
An analogous estimate for the   
neutron inside a nucleus gives the following 
life-time for $q<3$ (for $q=3$ and higher, the sum is divergent and
the contribution is enhanced by a coefficient $\sim n_{max}^{q-3}$)
\begin{equation}
\tau_n \simeq 10^{38}~(L{\rm GeV})^{- 3}~{\rm GeV}^{-1}~.
\end{equation}
One can set the bound on $L$ from the experiments
that look for the invisible decays of nucleons. The most sensitive
are the searches for $n \rightarrow \nu\nu\bar{\nu}$ channel\cite{kamio}.
This experiments look for the  $\gamma$-rays emitted in the
 deexcitation of the
nucleon hole produced by nucleon decay (or disappearance). 
The corresponding lower limit
on a partial lifetime is $\sim 5 \times 10^{26}$ years.
This implies that the period of an 
extra time must not exceed $L \sim 10^{12} ~M_{\rm P}^{-1}$.

 However, the following discussion suggests that the
applicability of such constraints is far less obvious.
Let us try to understand better the origin of the complexity.
For this it is useful to compute a one-loop correction to the
self-energy of an ordinary fermion (e.g. electron) due to KK tachyon exchange.
For simplicity, we shell consider a spin-1 tachyon of mass $\mu$.
One can do computation for real $\mu$ and then perform analytic continuation.
Using for instance the Pauli-Villars regularization,
one gets a well known result (for zero external momentum)
\begin{equation}
{\Delta m \over m}  \sim \int_0^1dx(2-x)log\left ({x\Lambda^2 \over
(x-1)^2m^2 + x\mu^2} \right)~.
\label{masscor}
\end{equation}
The logarithm has a branch cut for imaginary $\mu$, which gives complexity.
Formally this
corresponds to an electron decaying into electron and a tachyon
with a {\it negative} energy, 
$E_{tachion} = -\sqrt{{\overrightarrow p}^2 -\mu^2}$. 
Since this decay is in principle
allowed by the energy-momentum conservation, it creates a branch cut in
(\ref{masscor}).
Note the difference between the tachyon and a non-tachyonic boson
with negative energy $E = -\sqrt{{\overrightarrow p}^2 + \mu^2}$. The energy
conservation alone would be enough to prevent
a similar decay of the electron into such a boson with negative energy. 
However, it is not enough to prevent decay into the tachyon.
Thus, if the negative energy tachyons were ``real'',
the complexity in the self-energy 
could have been interpreted as the decay into these states.
However, existence of such states would contradict to the energy-positivity
condition. Thus, any sensible quantum theory of tachyons is expected in 
some way
to eliminate such states from the physical spectrum.
Perhaps most trivially, by imposing the energy
positivity condition and reinterpreting the negative energy tachyons
as the positive energy anti-tachyons.
We do not intend here to suggest any approach
towards building such theories. However, it is reasonable to think that
whatever mechanism eliminates negative and imaginary
(${\overrightarrow p}^2 < -\mu^2$) energy states from the physical spectrum,
must also render 
(\ref{masscor}) unphysical. If this is the case then the above
constraints from the nucleon decay should presumably be disregarded.

Note that in our estimates  it was crucial that we cut-off the number of
possible KK excitations by $n_{max}$. 
However, if one sums  
an infinite tower of KK excitations  the situation becomes more delicate. 
For instance,  
consider the case of a single extra time.
The potential mediated by the infinite sum of KK modes takes the form
\begin{equation}
V(r) \propto -i {\rm cot}\left ({r \over 2L}\right ){1 \over r} = 
-{i\over r} \left ( {2L \over r} + \sum_{k=1}^{\infty} {4rL \over
r^2 - 4k^2\pi^2L^2} \right )~.
\label{cot}
\end{equation}
Let us compute the gravitational self-energy of a  
spherical body of the 
radius $R$. This energy  is proportional to the 
integral of the potential (\ref {cot}) from zero to $R$. 
The answer depends on how many
poles of  (\ref {cot}) contribute to the integral.
In the simples case when all the poles of (\ref {cot})
are outside  of the volume of the body, i.e. $R<2\pi L$,
the expression for the energy is complex and does not have
a real part. This means that classical gravity is ``screened''
at very small distances by the corresponding KK states. 
In the case when  $R>2\pi k L$ (for some finite set of $k$'s)
the poles of (\ref {cot}) would produce a nonzero real part in the 
expression for the self-energy. 
This can be calculated by doing the integral explicitly and 
estimating the corresponding finite sum over the poles. 
The result takes the following form:
\begin{equation}
{{\rm Re}E\over M} \sim  G_N \rho \left [2 \pi RL + R^2 \right ]~.
\label{en}
\end{equation}
The second  term in this expression 
comes from the zero mode graviton that mediates a
normal Newtonian potential. The corrections due to KK modes, 
however, are only linearly
suppressed in $L$. 
Let us now turn to the imaginary part of the 
self-energy. It is defined by the regular part of 
(\ref {cot}) and can be calculated. Along with the standard terms
proportional to  either $iRL$ or $iR^2$ there appear terms
proportional to $iRL{\rm log}[(R-2\pi kL)/(R+2\pi k L)]$. 
These are singular as poles of (\ref {cot}) 
coincide with $R$, i.e. when $R=2\pi k L$.
This type  of singularities  could indicate that decoupling 
in the theory might not be happening, and that a possible  
right way to deal with 
this models is to consider them as effective theories with a cutoff. 
Even if we are away from the singularities mentioned above,
the model with an infinite number of KK states 
would put a much more severe constraint
on the size of extra times.
The reason for such a different behavior is the same as before. 
When an infinite number of KK states are taken into account the potential 
develops an infinite number of poles (\ref {cot}). Some 
of these contribute to the imaginary part of self-energy,
thus making ``decay'' width of the object much bigger. As a result
one needs to impose a stronger bound on the size of extra
time dimensions. 

In the estimates presented above  
$L$ was assumed to be smaller than the size of the body in question.
One might wonder about  the  cases when $R<<L$. 
To study this issue 
let us compute the
Newtonian potential between two point-like 
masses $m$ which are  localized at $\tau=0$  
at $r<<L$ distance apart. 
In the infinite volume approximation the potential 
can be  defined as follows:
\begin{equation}
V(r)t = {1 \over M_{Pf}^{2+q}}\int dx^{4 + q}dx^{'4 +
q}T^{AB}(x)~G_{AB,CD}(x-x')~T^{CD}(x')~,
\end{equation}
where $G_{AB,CD}$ is the graviton propagator, 
$T^{AB}(y) = (0,0,..0,m,0,0,0)\delta^q(\tau)\delta({\overrightarrow y}
-{\overrightarrow x})$ and $t$ is interaction time.
For the static case this gives the following expression:
\begin{equation}
 V(r) \sim  (i)^q{m^2 \over M_{Pf}^{2 + q}} {1\over r^{1 + q}}~.
\end{equation}
The unusual thing about this expression is that it is pure imaginary
for odd $q$, meaning that Newtonian potential is ``screened''
at $r<<L$ distances. For $q =1$ this can be directly read off
(\ref{cot}) by taking the $L \rightarrow \infty$ limit.

 Above should mean that the Newtonian gravity shuts-off at small distances.
The gravitation self-energy of a spherical body of size $R << L$
for finite $L$ can be directly computed from (\ref{cot}), since in this
case we do not encounter any poles
\begin{equation}
{E\over M} \sim iG_N \rho LR\left [1  + O({R \over L})^2 \right ]
\sim iG_{N(5)}\rho R\left [1  + O({R \over L})^2\right ]~.
\label{imgrav}
\end{equation}
where $G_{N(5)} \sim M_{Pf}^{-3}$ is a five-dimensional Newtonian constant.
Note that an analogous estimate for one extra space dimension would
give
\begin{equation}
{E\over M} \sim G_N \rho LR\left [1 + O({R \over L})^2 \right ]
\sim G_{N(5)}\rho R\left [1  + O({R \over L})^2 \right ]~.
\label{space}
\end{equation}
since in this case the analog of (\ref{cot}) is
\begin{equation}
V(r) \propto - {\rm coth}\left ({r \over 2L}\right ){1 \over r} = 
-{1\over r} \left ( 1 + 2\sum_{n=1}^{\infty}
{\rm exp} \left (-{rn \over L} \right ) \right )~.
\label{cth}
\end{equation}
If we cut-off the number of KK modes by $n_{max} \sim M_{Pf}L$, then
a nonzero real part will appear in eq. (\ref {imgrav})
\begin{equation}
{\rm Re}{E\over M} \sim G_N \rho {L\over M_{Pf}}
\label{imgrav1}
\end{equation}
This is suppressed by a factor $\sim (RM_{Pf})$ relative to 
(\ref{space}) as one naively could have guessed, since for
bodies as small as an inverse ultraviolet cut-off $M_{Pf}^{-1}$,
there are no KK modes available for ``screening'' gravity.
 How large $L$ can be in such a case? As we explained above, if
one disregards complexity in the potential as being unphysical,
the probability-violation bounds can be ignored. In such a case
the size of extra times can be within the reach of proposed
submillimeter measurements\cite{measurements}.
Thus for an odd number of extra times
these experiments should see the strength of the
Newtonian gravity diminishing at short distances!

However, if complexity may really signal the probability non-conservation,
$L$ presumably can not be in submillimeter range,
even in the case of even $q$
(in which case the potential is real for infinite $L$).
If complexity appears
in the amplitude, on the dimensional grounds,  
it should be of order one at distances $r \sim L$. 
Since the four- and high-dimensional
Newtonian laws must match at the size of extra 
dimensions, the potential at $r \sim L$
should have imaginary part comparable to the
(four-dimensional) gravitational strength
\begin{equation}
 {\rm Im} V \sim  {m^2 \over M_P^2}~ {1\over L}~.
\end{equation}
If regarded as probability violation,
for two neutrons separated by a distance 
$L$ this would imply  a disappearance rate
$\Gamma \sim 10^{-38}/(L{\rm GeV})$ GeV, 
which for $L \sim 1~{\rm mm}$ gives the lifetime
$t \sim 10^{51}~ {\rm GeV}^{-1}$ clearly contradicting with
an above derived bound. 


Finally, we would like to quote somewhat less severe bounds
which  may come from
the production of the tachyonic KK gravitons in the stars.
A single graviton production rate is  $\sim T(T/M_P)^2$. 
The total rate is enhanced by the multiplicity of
final states 
\begin{equation}
\sim T(T/M_P)^2(TL)^q~,
\end{equation}
where $T$ is a temperature in the star. 
Expressing this in terms of the fundamental Planck length
we get a suppression factor analogous to the one obtained in the 
case of extra spatial dimensions \cite{add}
\begin{equation}
\sim T(T/M_{Pf})^{2 + q}~.
\end{equation}
The two-graviton production rate is suppressed by 
extra powers of $M_P$
\begin{equation}
\sim T(T/M_P)^4(M_{Pf}L)^q \sim T^5/(M_{Pf}M_P)^2~,
\end{equation}
and is sub-dominant.

We would like to thank  B. Bajc, U. Cotti, G. Farrar, M. Porrati,
V. Rubakov, A. Sirlin, A. Vilenkin
and D. Zwanziger for discussions.
It is a pleasure to thank P. Huggins, 
A. Mincer, P. Nemethy  and E. Schucking 
for discussions on various astrophysical and
laboratory bounds on matter disappearance.

\end{document}